\documentclass[12pt]{article}
\usepackage{amsfonts}
\usepackage{latexsym}
\usepackage{amsmath,amssymb}
\usepackage{verbatim}
\usepackage{setspace}
\usepackage{cite}
\usepackage[textheight=9in, textwidth=6.5in, letterpaper]{geometry}

\numberwithin{equation}{section}
\usepackage{amsmath}
\usepackage{color}
\usepackage{hyperref}

\newcommand{\bea}{\begin{eqnarray}}
\newcommand{\eea}{\end{eqnarray}}
\newcommand{\be}{\begin{equation}}
\newcommand{\ee}{\end{equation}}

\makeatletter

\renewcommand\section{\@startsection {section}{1}{\z@}%
  {-3.5ex \@plus -1ex \@minus -.2ex}%
  {2.3ex \@plus.2ex}%
  {\normalfont\large\bfseries}}

\renewcommand\subsection{\@startsection{subsection}{2}{\z@}%
  {-3.25ex\@plus -1ex \@minus -.2ex}%
  {1.5ex \@plus .2ex}%
  {\normalfont\bfseries}}

\begin{document}

\title{ \LARGE {\textsc {Central charges for AdS black holes}}}

\author{
{\large Malcolm Perry}$^{1,2,3}$\footnote{malcolm@damtp.cam.ac.uk.}, {\large Maria J. Rodriguez}$^{4,5}$\footnote{majo.rodriguez.b@gmail.com}, \\
\\
  $^{1}${\small Department of Physics, Queen Mary University of London,}\\{\small Mile End Road, London E1 4NS, UK}\\
\\ 
  $^{2}${\small DAMTP, Cambridge University, Centre for Mathematical Sciences,}\\ {\small  Wilberforce Road, Cambridge CB3 0WA, UK}\\
  \\
  $^{3}${\small Trinity College, Cambridge, CB2 1TQ, UK}\\
  \\
  $^{4}${\small  Department of Physics, Utah State University,}\\ {\small 4415 Old Main Hill Road, UT 84322, USA}\\
    \\
  $^{5}${\small  Instituto de F\' isica Te\' orica UAM/CSIC, Universidad Aut\'  onoma de Madrid,}\\ {\small  13-15 Calle Nicol\' as Cabrera, 28049 Madrid, Spain}\\
 \\
\\
 }
\maketitle
\abstract
Nontrivial diffeomorphisms act on the horizon of a generic 4D black holes and create distinguishing features referred to as soft hair. Amongst these are a left-right pair of Virasoro algebras with associated charges that reproduce the Bekenstein-Hawking entropy for Kerr black holes. In this paper we show that if one adds a negative cosmological constant, there is a similar set of infinitesimal diffeomorphisms that act non-trivially on the horizon. 
The algebra of these diffeomorphisms gives rise to a central charge. Adding a boundary counterterm, justified to achieve integrability, leads to well-defined central charges with $c_L = c_R$. The macroscopic area law for Kerr-AdS black holes follows from the assumption of a Cardy formula governing the black hole microstates.

\newpage

\tableofcontents

\section{Introduction}
\label{sec:Intro}

The entropy of black hole can be understood by a type of 'soft hair' associated to a general class of nonabelian Vir$_L$ $\times$ Vir$_R$ diffeomorphisms. In fact, only recently, a general class of Vir$_L$ $\times$ Vir$_R$  diffeos of a generic spin $J$ Kerr black hole were considered in \cite{Haco:2018ske} to determine its entropy in a manner similar to their stringy black hole counterparts.

\indent  Strikingly, using the horizon itself as the surface permitting the analysis allow one to infer that there are no independent interior black hole microstates. In the case of black hole solutions in General Relativity, the black hole Hilbert space must be contained within the Hilbert space of states on or outside the black hole horizon. The observations seem to apply equally to the real-world Kerr black hole case \cite{Haco:2018ske}, the Kerr-Newman black hole solution \cite{Haco:2019ggi} and to the stringy black holes with near-AdS$_3$ regions.

\indent An interesting property of many non-supersymmetric black holes, such as Kerr and Kerr-AdS, is that  its entropy remains finite for the extreme configuration where the Hawking temperature vanishes. This suggests that for extremal black holes the entropy can be reproduced as the statistical entropy of the dual CFT using the Cardy formula.
Previous attempts to determine the microscopic entropy of the black hole and reproduce the macroscopic area law relied on the existence of a near horizon extremal-AdS$_2$ region. The so called Kerr/CFT correspondence \cite{Guica:2008mu}, employed the near horizon extremal  (zero-temperature) Kerr to show that the microscopic entropy of the CFT, calculated using the Cardy formula, coincides with the macroscopic the extremal Kerr black hole entropy.  \cite{Castro:2010fd} generalized these ideas, only assuming that the left and right CFTs have identical central charges, and argued that even away from extremality Kerr black hole's entropy was reproduced. Similar results were presented for the five-dimensional Myers-Perry black holes in \cite{Krishnan:2010pv} and non-vanishing cosmological constant  in \cite{Mao:2016pwq,Grumiller:2016kcp}.

\indent  In this paper we follow the soft-hair approach in a slightly different direction. In an effort to test the applicability of the soft-hair conjecture we will focus on the Kerr-AdS black holes. The twofold goal is to uniquely define the conformal phase space formalism relevant to the approach and find a boundary counterterm to achieve integrability for the existence of a well-defined charges. Assuming the existence of a quantum Hilbert space on which these charges generate the symmetries, as well as the applicability of the Cardy formula, the concrete example of Kerr-AdS that we explore gives identical central charges $c_L = c_R$ providing evidence for an actual explanation of the macroscopic entropy.

\indent The existence of a generic near-AdS$_3$ regions is not ordinarily sufficient to determine the conformal coordinates in which the Virasoro action takes a simple form and exhibits the conformal structure of the black hole geometry. To supplement the approach, we will argue that the Noether charge interpretation of the entropy first presented in \cite{Wald:1993nt} gives the necessary input to uniquely fix the relevant conformal coordinates. We will only assume that the theory admits stationary black hole solutions with a bifurcate Killing horizon.

\indent  Having established the appropriate conformal coordinates, we will construct an explicit set of Vir$_L$ $\times$ Vir$_R$ vector fields which generate the hidden conformal symmetry in the near-horizon region. These act non-trivially on the horizon in the sense that their boundary charges are non-vanishing. The covariant phase space formalism provides a formula for the Virasoro charges as surface integrals on the horizon. Using the covariant phase space formalism we find a boundary counterterm -- that reduces to the Wald-Zoupas boundary counterterm for Kerr -- justified to achieve integrability for well-defined charge and gives $c_L = c_R$.

The plan of the paper is as follows. In Section 2 we use the entropy as Noether charge interpretation \cite{Wald:1993nt} to find expressions for the generators of these $R,L$ sectors in a {\it conformal coordinate} frame  with a direct CFT interpretation. Section 3 presents the identification of the AdS$_3$ near horizon in conformal coordinates for generic Kerr-AdS black holes in which the Virasoro action takes the simple form. The application of the blended AdS$_3$ near horizon identification and $R,L$ entropy killing field generators relevant for its Noether construction is presented in Section 4. This section includes the explicit identification of the microscopic temperatures $T_L$ and $T_R$ in Kerr-AdS. In Section 5 we compute the covariant right-left moving Virasoro charges and find a counterterm for an integrable action. We then indicate how to do a similar calculation for the left-moving charges. In section 6 we discuss the case of extreme black holes. In Section 7 we briefly discuss the first law.  Section 8 repeats the same exercise but for the inner black hole horizon. The discussion of our findings are provided in Section 9. We illustrate this black hole conformal structure method with the detailed Kerr black hole example in Appendix A.\\

Throughout this paper we use units such that $c=\hbar=k=G=1$.

\section{Entropy and $(L,R)$\,-\,generators}
\label{sec:Scoords}

Consider a stationary black hole solution with a bifurcate Killing horizon. According to Wald's computation \cite{Wald:1993nt}, the black hole entropy $S_{\pm}$ is simply $2\pi$ times the integral over the Noether charge associated with the Killing field 
\bea \label{killing}
\bar{\zeta}_{\pm}=\kappa_{\pm}\,{\zeta}_{\pm}\,,
\eea
Here $\kappa_{\pm}$ is the surface gravity \footnote{The Hawking temperature is given by $T_{\pm}=\kappa_{\pm}/(2\pi)$.} and ${\zeta}_{\pm}=\partial_t+\sum_i \, \Omega_{\pm}\partial_{\phi}$ are the horizon Killing fields (normalized so as to have unit surface gravity) vanishing on the bifurcation  $2$-surfaces $\Sigma_{\pm}$. With the aim to find a frame with a direct CFT interpretation for the outer/inner event horizons, we proposed to consider new variables $t_{L,R}$ where the Killing field (\ref{killing}) generating the entropy is
\bea \label{prop}
\bar{\zeta}_{\pm}\propto (\partial_{t_L}\pm\partial_{t_R})\,.
\eea
and vanishes respectively either on the inner or outer event horizon. This is possible, for $4$-dimensional black holes via the transformation
 \bea\label{trtl}
t_R&=& \alpha \, \phi+\beta\, t \,, \\
t_L&=& \gamma \,\phi+\delta \,t \,.\nonumber
\eea
for the following choice of parameters
\bea \label{trtl1}
&&\beta=-\frac{1}{2}   \left[\gamma \left(\Omega_+ - \Omega_-\right) + \alpha \left(\Omega_+ + \Omega_-\right)\right]\,\\
&&\delta=-\frac{1}{2}   \left[\alpha \left(\Omega_+ - \Omega_-\right) + \gamma \left(\Omega_+ + \Omega_-\right)\right]\,.\nonumber
\eea
This choice, gives the Killing fields (\ref{killing}) in a frame where the right (R) and left (L) sectors of the CFT is straightforward
\bea \label{zeta}
\bar{\zeta}_{\pm}=\pm \kappa_{\pm} K_{\pm}  (\partial_{t_L}\mp \partial_{t_R})\,.
\eea
where
\bea
K_{\pm}=\frac{1}{2} \left(\gamma \mp \alpha \right) (\Omega_- -\Omega_+)\,.
\eea
We can also now invert the relations to find expressions for the generators of these $R,L$ sectors \footnote{In turn, we can also rewrite the above formulae in the form
\bea
\partial_{t_L}=\frac{1}{2} \left[\left(\frac{1}{K_+}-\frac{1}{K_-}\right)\partial_t+\sum_i \left(\frac{\Omega_+^{(\phi_i)}}{K_+}-\frac{\Omega_-^{(\phi_i)}}{K_-}\right)\partial \phi_i\right]\,,\\
\partial_{t_R}=\frac{1}{2} \left[\left(\frac{1}{K_+}+\frac{1}{K_-}\right)\partial_t+\sum_i \left(\frac{\Omega_+^{(\phi_i)}}{K_+}+\frac{\Omega_-^{(\phi_i)}}{K_-}\right)\partial \phi_i\right]\,,
\eea
}.
The choice of $t_R,t_L$ coordinates that we have derived is very suggestive of the fact that there maybe a universal correspondence of a 2d CFT for any 4-dimensional black hole solution. To realize this conjecture we derive, in the following section, a method to identify the microscopic temperatures $T_L$ and $T_R$ (or equivalently the parameters $\alpha, \gamma$). While the conformal coordinates are not uniquely defined, the choice made here is unique in the sense that it would later give the correct thermodynamic relations that verify of the first law of thermodynamics for black holes. As we will now show the coordinate choice $(t_L,t_R)$ also matches those found by computing the monodromies around the inner and outer horizon in \cite{Castro:2013kea}.

\section{Conformal coordinates}
\label{sec:ConfCoords}

In this section we propose a method for the identification of the microscopical temperatures of 2d CFT that relies on the properties of the black hole geometry close to the bifurcation surfaces. This work is a logical outgrowth of the observation in \cite{Haco:2018ske} that the existence of a generic near-AdS$_3$ regions constrains the define the so-called conformal coordinates $(w^\pm,y)$ and possibly the structure of the black hole geometry. 

There is a large ambiguity in the choice of the conformal coordinates in which the Virasoro action takes the simple form. The procedure does not fix uniquely the 2d CFT temperatures $T_{L,R}$. For the purpose of CFT identification, one can take Wald's  construction to define the black hole entropy as a Noether charge associated with the Killing fields in a frame where the right (R) and left (L) are distinguishable (see Section \ref{sec:Scoords} for details). This suggests a natural modification of the warped AdS space-times geometry identification. We propose the coordinate choice to be consistent with (\ref{trtl})-(\ref{trtl1}) where we have decomposed the coordinates, which separate the corresponding Killing fields in a (R,L) frame.

We now conjecture that the conformal coordinates $(w^\pm,y)$ clearly exhibiting the conformal structure are
\bea\label{CFT}
w^+&=&R(r)\,e^{t_R}\,,\nonumber\\
w^- &=&R(r)\,e^{t_L}\,\\
y&=&Q(r)\,e^{(t_L+t_R)/2}\,\nonumber
\eea
where $(t_L,t_R)$ are defined in (\ref{trtl}) with (\ref{trtl1}), $R(r)^2+Q(r)^2=1$ and close to the event horizon $r=r_+$ the function $R^2(r) \sim c \,(r- r_+)$ with constant $c$. The bifurcation surface in the conformal coordinates at the outer event horizon is at $w_{\pm}=0$ and $w_{\pm}=\infty$ at the inner event horizon. 

It follows from (\ref{CFT}) that a black hole solution (in Boyer-Lindquist type coordinates) under the transformation
\bea\label{eq:coordtransf}
t&=& \frac{1}{2(\gamma \beta- \alpha \delta)}\left[(\alpha+\gamma) \ln\frac{w^+}{w^-}-(\alpha-\gamma) \ln{(w^+w^-+y^2)}\right]\,,\nonumber\\
\left(\frac{R(r)}{Q(r)}\right)^2&=&\frac{w^+ w^-}{y^2}\,\\
\phi&=&\frac{1}{2\alpha} \, \ln \frac{w^+ (w^+w^-+y^2)}{w^-} -\frac{\beta}{\alpha} \, t\,,\nonumber
\eea
becomes at the leading order around the bifurcation surface
\bea
ds^2= \frac{4\,\rho_+(\theta)^2}{y^2}dw^+ dw^-+\frac{F(\theta)\sin^2\theta}{y^2 \rho_+(\theta)^2}dy^2+\rho_+(\theta)^2d\theta^2+...\,,
\eea
for specific values of $\alpha$ and $\gamma$. Here $\rho_+(\theta),F(\theta)$ are arbitrary functions of $\theta$. This is simply a warped AdS spacetime in conformal coordinates. Hence these coordinates are well-adapted to an analysis of 4D black holes mirroring that of the 3D BTZ black holes.

Physically this means that there is a unique  2d CFT dual description for generic types black hole geometries. We observe that under azimuthal identification $\phi\rightarrow\phi+2\pi$  of the conformal variables
\bea
w^+\sim e^{2\pi \alpha} w^+\,,\qquad w^-\sim e^{2\pi \gamma} w^-\,,\qquad y \sim e^{\pi(\alpha+\gamma)} y\,.
\eea
This is the same as the identification employed in \cite{Maldacena:1998bw} that turns $AdS_3$ in Poincare coordinates into BTZ
with temperatures $(T_L, T_R)$
\bea
w^+\sim e^{4\pi^2T_R} w^+\,,\qquad w^-\sim e^{4\pi^2 T_L} w^-\,,\qquad y \sim e^{2\pi^2(T_R+T_L)} y\,.
\eea
The periodicities analysis yields
\bea
\alpha=2\pi T_R\,,\qquad \gamma=2\pi T_L\,.
\eea
or equivalently
\bea\label{temRL}
T_R=\alpha/(2\pi)\,,\qquad T_L=\gamma/(2\pi)\,.
\eea
We argue, that with the proposed systematic approach one can identify a CFT dual for generic types black hole geometries, allowing the computation of
the left and right $2d$ CFT temperatures and the microscopic entropy using a Cardy formula. In the next section, we present results for the $2d$ CFT temperatures in explicit examples that include Kerr black hole, and Kerr-AdS black holes that are the main focus of the paper. 

\section{Conformal Coordinates for Kerr-AdS Black Hole}

In the previous section we made a rather general proposal to identify the 2d CFT temperatures via the conformal coordinates definitions. In this section we put flesh on this proposal and show the explicit application of the proposed method to the Kerr-AdS black hole solution. In the flat space-time limit $L\rightarrow \infty$ our results reduce to the Kerr black hole identifications found in \cite{Haco:2018ske} (see the appendix).

The metric of the four dimensional Kerr-AdS black hole  \cite{Carter:1968ks}, satisfying $R_{\mu\nu}=-3L^{-2}\,g_{\mu\nu}$ is given by
\bea
ds^2&=&\rho^2\left(\frac{dr^2}{\Delta}+\frac{d\theta^2}{\Delta_{\theta}}\right)+\frac{\Delta_{\theta}\sin^2\theta}{\rho^2}\left(a\,dt-\frac{r^2+a^2}{\Xi}d\phi\right)^2-\frac{\Delta}{\rho^2}\left(dt-\frac{a\sin^2\theta}{\Xi} d\phi\right)^2\,,\nonumber\\
\rho^2&=&r^2+a^2\cos^2\theta\,\qquad \Delta=(r^2+a^2)(1+r^2L^{-2})-2Mr\,,\nonumber \\
&&\Delta_{\theta}=1-a^2L^{-2}\cos^2\theta\,,\qquad \Xi=1-a^2L^{-2}\,.
\eea
The metric is asymptotic to $AdS_4$ in a rotating frame, with angular velocity $\Omega_{\Lambda}=-a L^{-2}$. The outer and inner event horizons are located at $\Delta(r_{\pm})=0$. The physical parameters are
\bea\label{thermodynamicsAdS}
E=\frac{M}{\Xi^2}\,,\qquad J&=&\frac{M a }{\Xi^2}\,,\\
T_{\pm}=\frac{\Delta'(r_{\pm})}{4\pi(r_{\pm}^2+a^2)}\,\qquad 
 \Omega_{\pm}&=&\frac{a\,\Xi}{r_\pm^2+a^2}\,,\qquad S_{\pm}=\frac{\pi (r_{\pm}^2+a^2)}{\Xi} \,.\nonumber
\eea
corresponding to the physical mass, angular momentum, Hawking's temperature, angular velocity of the horizon  (as measured in the asymptotically rotating frame), and the entropy respectively. As shown in \cite{Gibbons:2004ai}, the angular velocity which is measured relative to a {\it non-rotating} frame at infinity is determined by
\bea
\Omega^{\pm}_{\infty}=\Omega_{\pm}-\Omega_{\Lambda}
=\frac{a (1+r_\pm^2/L^2)}{r_\pm^2+a^2}\,.
\eea
To proceed with the interpretation, we first take the transformation (\ref{eq:coordtransf}) with (\ref{trtl1}), and set $c = 1/\Delta'(r_+)$ in the function $R^2(r) \sim c \,(r- r_+)$ close to the event horizon of the black hole, such that the black hole metric around the bifurcation surface is 
\bea
ds^2&=& \frac{4\,\rho_+(\theta)^2}{y^2}dw^+ dw^-+\frac{k^2\, \Delta_\theta(\theta) \sin^2\theta}{y^2 \rho_+(\theta)^2}dy^2+\frac{\rho_+(\theta)^2}{\Delta_\theta(\theta)}d\theta^2\nonumber \\
&+& \frac{4 w^{+}}{y^3} \left(\frac{a^2 \Delta_{\theta}(\theta)\sin ^2 \theta  \,( 2\, \delta \, r_+  +1) }{\delta ^2   \rho_+(\theta)^2 }+\frac{\left(a^2+r_+^2\right) (r_- -r_+)} {r_+} \right) dw^{-} dy\\
&-& \frac{4 w^{-}}{ y^3  } \left(\frac{a^2 \Delta_{\theta}(\theta) \sin ^2(\theta ) \, (2 \,\delta \, r_+ -1) } {\delta ^2   \rho_+(\theta)^2}+\frac{(a^2+r_+^2) (r_- +r_+ )-2 a^2 r_+ \sin^2 \theta }{r_+} \right)dw^{+} dy\nonumber \\
&+&...\,,\nonumber
\eea
where 
\bea
\rho_+(\theta)^2=r_+^2+a^2\cos^2\theta\,,\qquad k= 2 a /\delta\,.
\eea
In this case, we find the parameters in the transformation become 
\bea
\alpha=\frac{\Delta'(r_+)}{2 \,a\Xi}\,,\qquad \beta=0\,,\qquad\gamma=\frac{(\Omega_+ +\Omega_-)}{(\Omega_- -\Omega_+)} \alpha
\,,\qquad \delta=-\Omega_+ (\alpha+\gamma)
\eea
and from (\ref{temRL}) one can identify the right and left-temperatures arising from the CFT giving
\bea\label{AdsTemp}
T_R=\frac{\Delta'(r_+)}{4\pi \,a\Xi}\,,\qquad T_L=\frac{(\Omega_++\Omega_-)}{(\Omega_- -\Omega_+)}\frac{\Delta'(r_+)}{4\pi \,a\Xi }\,.
\eea
%
In the case of vanishing cosmological constant, $L\rightarrow \infty$, the results for the CFT temperatures reduce to those for Kerr black holes given in \cite{Haco:2018ske}. Further details about the conformal coordinates for Kerr are presented in Appendix \ref{app:Kerr}.

The general result (\ref{AdsTemp}) for the right and left-temperatures arising from the CFT for the AdS-Kerr black holes can be combined with the Cardy entropy formula 
\begin{equation}
S_{\pm}=\frac{\pi^2}{3}\left(c_L T_{L}\pm c_R T_R \right)\,.
\end{equation}
Using (\ref{AdsTemp}) and the entropies (\ref{thermodynamicsAdS}), one finds that for Kerr-AdS black holes the left and right CFTs have identical central charges
\bea\label{centralAdsKerr}
c_L=c_R=-\frac{6 a  }{\delta }=\frac{ 6 a (r_+^2-r_-^2)}{\Delta'(r_+)}\,.
\eea
%

\subsection*{Inverse metric}
\begin{eqnarray}
&& g^{yy}\sim \frac{\delta ^2 y^2  \rho_+^2}{4 a^2 \sin^2\theta \Delta_{\theta} (\theta )}\,,\qquad
 g^{\theta\theta}\sim\frac{\Delta_{\theta}}{\rho_+^2}\,,\qquad
g^{+ -} \sim \frac{y^2}{4 \rho_+^2}\,,\nonumber\\
&& g^{+ y} \sim- \frac{w_+ y}{4}\left(\frac{2 \, \delta  \, r_+ +1}{\rho_+^2}- \frac{\delta ^2 \left(a^2+r_+^2\right) (r_+ - r_-)}{a^2 r_+ \sin^2\theta\, \Delta_{\theta}}\right)\,,\\
&& g^{- y} \sim \frac{ w^{-} y}{4  } \left(\frac{2 \,\delta \, r_+ -1 } { \rho_+^2}+\frac{\delta ^2 ((a^2+r_+^2) (r_- +r_+ )-2 a^2 r_+ \sin^2 \theta) }{a^2 r_+ \sin^2\theta\, \Delta_{\theta}} \right)\,.\nonumber
\end{eqnarray}

\subsection*{Volume element}
The volume element is
\bea
\epsilon_{+- y \theta} =\frac{4 a \sin\theta  \rho_+^2 }{\delta y^3}+ ... \,.
\eea

\subsection*{Conformal vectors}
We consider the conformal vector fields
\bea\label{eq:vectorfields}
\zeta_{n}&=& \epsilon_n \partial_+ +\frac{1}{2} \partial_+ \epsilon_n y \partial_y \,,\\
\bar{\zeta}_{n}&=&  \bar\epsilon_n \partial_- +\frac{1}{2} \partial_- \bar\epsilon_n y \partial_y \,.
\eea
and restrict the full set of functions $(\epsilon ,\bar\epsilon)$ so that $(\zeta, \bar\zeta)$ are invariant under $2\pi$ azimuthal rotations is
\bea
\epsilon_n &=& \alpha \,(w^+)^{1+\frac{i n}{\alpha}}=2\pi T_R(w^+)^{1+\frac{i n}{2\pi T_R}}\,, \\
\bar\epsilon_n  &=& \gamma \,(w^-)^{1+\frac{i n}{\gamma}}= 2\pi T_L(w^-)^{1+\frac{i n}{2\pi T_L}}\,.
\eea
Taking $\zeta_n\equiv \zeta(\epsilon_{n})$ and  $\bar\zeta_n=\bar\zeta(\epsilon_{n})$, the vector fields (\ref{eq:vectorfields}) obey the Lie bracket algebra
\bea
[ \zeta_m,\zeta_n ] = i (n-m)\zeta_{m+n} \,, \qquad
[ \bar\zeta_m, \bar \zeta_n ] =  i (n-m) \bar\zeta_{m+n}\,.
\eea
and the two set commuting with another
\bea
[\zeta_m,\bar\zeta_n]=0\,.
\eea
The zero modes in this case are
\bea
\zeta_0&=& \alpha (w^+\partial_++\frac{1}{2} y\, \partial_y)=2\pi T_R(w^+\partial_++\frac{1}{2} y\, \partial_y)= \partial_{\phi}-\frac{2\pi T_L}{\delta}\, \partial_t \equiv - i  \, \omega_R\,,\\
\bar \zeta_0 &=& \gamma (w^-\partial_-+\frac{1}{2} y \,\partial_y)= \frac{2\pi T_L}{\delta}\, \partial_t \equiv  i \, \omega_L
\eea


\section{Covariant charges}
\label{sect:CovCharges}

In this section we construct the linearized covariant charges $\delta \mathcal{Q}_n=\delta \mathcal{Q}(\zeta,h;g)$ associated to the diffeos $\zeta_n$ acting on the horizon. We are interested in the central term $K_{m,n}$ in the Virasoro charge algebra (for the right movers)
\bea
\{\mathcal{Q}_n,\mathcal{Q}_m\}=(m-n) \mathcal{Q}_{m+n}+K_{m,n}\,,
\eea
where the central term is given by
\bea
K_{m,n}=\delta\mathcal{Q}(\zeta_n,\mathcal{L}_{\zeta_m} g;g) =\frac{c_R \,m^3}{12} \delta_{m+n}\,.
\eea

It turns out that the general form for the linearized charge associated to a diffeo $\zeta$ on a surface $\Sigma$ with boundary $\partial \Sigma$ is
\bea
\delta \mathcal{Q} = \delta \mathcal{Q}_{IW}+\delta \mathcal{Q}_{ct}
\eea
On the one hand, there is a nonzero contribution from the Iyer-Wald charge 
\bea
\delta \mathcal{Q}_{IW}=\frac{1}{16\pi} \int_{\partial\Sigma} *F_{(IW)}
\eea
with
\bea
{F_{(IW)}}_{ab}=\frac{1}{2} \nabla_{a} \zeta_b h+\nabla_{a}{h^c}_b\zeta_c+\nabla_c \zeta_a {h^c}_b+\nabla_c {h^c}_a \, \zeta_b-\nabla_a h\, \zeta_b - a \leftrightarrow b\,.
\eea
Here we follow the conventions of \cite{Haco:2018ske}. The variation $h^{ab}$ is defined as $g^{ab}\rightarrow g^{ab}+h^{ab}$. We take the metric perturbation $h^{ab}=\mathcal{L}_{\tilde{\zeta}_m}g^{ab}$ due to the second diffeomorphism  $\tilde{\zeta}$ and  $h=h^{ab}g_{ab}$.\\


On the other hand, one also needs to add a counterterm
\bea
\delta \mathcal{Q}_{ct} =\frac{1}{16\pi}\int_{\partial\Sigma} {F_{(ct)}}_{ab}\, d \Sigma^{ab}\,
\eea
where $N$ is the volume two-form on the normal bundle to the $\Sigma_{bif}$. 
\bea
{F_{(ct)}}_{ab}=- 2 {N_{d}}^c\nabla_c(\zeta_a {h^d}_{b}) - a \leftrightarrow b\,,
\eea
Note that the addition of this counterterm is justified to achieve integrability. The nonzero contributions to $K_{n,m}$ come only from 
\bea
\delta\mathcal{Q}=\frac{1}{16 \pi} \int d\theta dw^+ \epsilon_{\theta + - y} ({F_{(IW)}}^{-y}+{F_{(ct)}}^{-y})
\eea
We find that
\bea
{F_{(IW)}}^{-y}=4 h^{y-}\zeta^y\Gamma^-_{y-}\,,
\eea
 Considering ${N_+}^+=1$, ${N_-}^-=-1$ then
 \bea
 {F_{(ct)}}^{-y}&=&-2 \nabla_+(\zeta^- h ^{+y})+2 \nabla_+(\zeta^y h ^{+-})+2 \nabla_-(\zeta^- h ^{-y})-2 \nabla_-(\zeta^y h ^{--})\\
 &=&2 \zeta^y(\nabla_+ h^{+-})+2 (\nabla_-\zeta^-) h^{-y}-2\zeta^y (\nabla_- h^{--})\\
 &=&2 \zeta^y h^{-y}(\Gamma^+_{+y}+\Gamma^-_{-y}-2 \Gamma^-_{-y})\\
 &=& 2 \zeta^y h^{-y} (\Gamma^+_{+y}-\Gamma^-_{-y})
 \eea 
Note that
\bea
h^{+-}=0\,, \qquad h^{--}=0\,,\qquad  \nabla_+  \zeta^-=0\,,\qquad \nabla_-  \zeta^-=\Gamma^-_{-y}\zeta^y\,,
\eea
and also
\bea
\nabla_+h^{+y}=0\,,\qquad \nabla_+h^{+-}=\Gamma^+_{+y} h^{-y}\,,\qquad \nabla_-h^{-y}=0\,,\qquad \nabla_-h^{--}=2\Gamma^-_{-y} h^{-y}\,.
\eea
Adding the terms together one finds
\bea
\delta\mathcal{Q}&=&\frac{1}{16\pi}\int d\theta dw^+  \epsilon_{+- y \theta} (4 h^{y-}\zeta^y\Gamma^-_{y-}+ 2 \zeta^y h^{-y} (\Gamma^+_{+y}-\Gamma^-_{-y}))\,,\\
&=&\frac{1}{16\pi}\int d\theta dw^+ \frac{4 a \sin\theta  \rho_+^2 }{\delta y^3}  2 h^{y-}\zeta^y(\Gamma^-_{y-}+\Gamma^+_{+y})\,,
\eea
By working at small $w^+$ and taking the $w^+$ limit (which amounts to approaching $\Sigma_{bif}$ along the future horizon) one finds
\bea 
&& h^{-y}=g^{+-}\partial_+ \zeta^y = \frac{y^3\tilde{\epsilon}''}{4\rho^2_+} \qquad \text{with} \qquad  '=\partial_+\,,\\
&& \int d\theta \sin\theta =2, \,\, \text{and} \qquad \Gamma^+_{+y}+\Gamma^-_{-y}=-\frac{2}{y}.
\eea
Choosing $\zeta$ to be $\zeta_n$ and $\tilde\zeta$ to be $\zeta_m$, the variation becomes
\bea
K_{m,n} = \delta\mathcal{Q}&=&\frac{1}{16\pi}\int d\theta dw^+ \frac{4 a \sin\theta  \rho_+^2 }{\delta y^3}  2 \left(\frac{y^3\epsilon_m''}{4\rho^2_+}\right) \left(\frac{1}{2} y \epsilon_n' \right)\left(\frac{-2}{y}\right) \,\\
&=&-\frac{1}{16\pi}\int d\theta dw^+ \frac{2 a \sin\theta  }{\delta }   {\epsilon_m''} \epsilon_n'  \\
&=&-\frac{1}{16\pi}\int  \frac{dw^+ }{w^+ }\frac{ 4 a  }{\delta }   {\epsilon_m''} \epsilon_n' \,,\\
&=&-\frac{1}{16\pi}(4\pi^2 T_R)\frac{ 4 a  }{\delta }   \frac{i m^3}{2\pi T_R} \delta_{m+n,0}\,\\
&=&\frac{ a  }{2\delta } i m^3 \delta_{m+n,0}\,.
\eea
Here we have computed the Dirac bracket of two charges. Passing to the commutator introduces a factor of $-i$
resulting in a central charge of
\bea
c_R=\frac{6 a  }{\delta }
\eea
in agreement with (\ref{centralAdsKerr}).

\subsection*{Left-moving charges}

The computation of the central charge in the left-moving sector exactly parallels the computation in the right moving sector. All one needs to do is to exchange $\epsilon$ for $\bar\epsilon$, interchange $w^+ \leftrightarrow w^-$ together with a similar exchange in the components of the tensors and Christoffel symbols and finally
replace $T_R$ by $T_L$. The result for the left-moving central charge $c_L$ is
\be c_L = \frac{6a}{\delta},\ee
which is exactly the same as $c_R$.   

\section{Extremal AdS-Kerr black hole}
In the extreme limit (when $T_{\pm} =0 $), the inner and outer horizons degenerate to a single horizon at $r_0\equiv r_+=r_-$.
The extremality condition is $\Delta(r_0)=\Delta'(r_0)=0$ that implies
\bea
a^2=\frac{r_0^2(1+3r_0^2/L^2)}{(1-r_0^2/L^2)}\,\nonumber \\
M=\frac{r_0(1+r_0^2/L^2)^2}{(1-r_0^2/L^2)}\,.
\eea

At extremality, the right temperature arising for the CFT vanishes. And, the left temperature in the extremal limit agrees exactly with previous results \cite{Hartman:2008pb,Lu:2008jk} 
\bea\label{tempextremal}
T_R=0\,,\qquad T_L=\frac{1+6r_0^2L^{-2}-3r_0^4L^{-4}}{2\pi(1-3r_0^2L^{-2})\sqrt{(1+3r_0^2L^{-2})(1-r_0^2L^{-2})}}\,.
\eea
where $r_0\le L/\sqrt{3}$ to restrict the values of $0<T_L$.
The central charge (\ref{centralAdsKerr}) becomes 
\bea\label{centralextremal}
c_L=\frac{12 r_0\sqrt{r_0^2(1+3 r_0^2/L^2)(1-r_0^2/L^2)}}{1+6 r_0^2/L^2-3r_0^4/L^4}\,.
\eea
For a unitary conformal field theory at temperature $T_L$, the microscopic entropy from the Cardy formula is given by
\bea
S_+=\frac{\pi^2}{3} c_L T_L\,.
\eea
From (\ref{centralextremal}) and (\ref{tempextremal}), we therefore obtain the microscopic entropy
\bea
S_+=\frac{2\pi r_0^2}{1-3r_0^2/L^2}\,.
\eea
This is in perfect agreement with the extreme black hole entropy given in (\ref{thermodynamicsAdS}). For vanishing cosmological constant $(L\rightarrow\infty)$ the results reduce to the asymptotically flat Kerr black hole.
%
%

\section{First Law of Thermodynamics}
In this section we discuss the first law of thermodynamics for the AdS-Kerr black holes \cite{Gibbons:2004ai}, that is given by
\bea
\delta E = \pm T_{\pm}  \delta S_{\pm} + \Omega^{\pm}_{\infty} \delta J\,.
\eea
Remarkably, the first law may also be written
\bea
\delta S_{\pm} = \frac{\delta E_L}{T_L} \pm \frac{\delta E_R}{T_R}\,.
\eea
where \footnote{As described in \cite{Castro:2013kea}, we consider the change $\phi\rightarrow \hat{\phi}- a\, t/L^2$ in (\ref{trtl}).}
\bea
\delta E_L&=&-\frac{\gamma}{\delta} \,\delta E+\frac{\gamma}{\delta} \frac{a }{  L^2} \, \delta J\,,\\
\delta E_R&=&-\frac{\gamma}{\delta} \, \delta E+ \left(-1+\frac{\gamma}{\delta} \frac{a }{ L^2}\right)\delta J\,,
\eea
and the right and left temperatures are defined in (\ref{AdsTemp}). We have also used here $\delta E_L=-\gamma\, \delta \mathcal{E}_L$  and $\delta E_R=\alpha\, \delta \mathcal{E}_R$ \footnote{Provided $\epsilon$ in the vector field $\zeta=\epsilon \partial_+ +\frac{1}{2}  \partial_+ \epsilon y  \partial_y$, and similarly $\bar{\epsilon}$ in $\bar{\zeta}=\bar{\epsilon} \partial_- +\frac{1}{2}  \partial_- \bar{\epsilon} y  \partial_y$, is invariant under $2 \pi$ azimuthal rotations \cite{Haco:2018ske} we define the normalized functions $\epsilon=2\pi T_{R} (w^+)^{1+\frac{i n}{2\pi T_R}}=\alpha \,(w^+)^{1+\frac{i n}{\alpha}}$ and $\bar{\epsilon}=2\pi T_{L} (w^-)^{1+\frac{i n}{2\pi T_L}}=\gamma \,(w^-)^{1+\frac{i n}{\gamma}}$.}.

\section{Inner black hole horizon}

We could choose the inner horizon $r_-$ instead to proceed with the approach. We first take the transformation (\ref{eq:coordtransf}) with (\ref{trtl1}), and set $\tilde{c}^2 = 1/\Delta'(r_-)$ in the function $R^2(r) \sim \tilde{c} \,(r- r_-)$ close to the Cauchy horizon of the black hole, such that the leading order of the black hole metric around the bifurcation surface is 
\bea
ds^2&=& \frac{4\,\rho_-(\theta)^2}{y^2}dw^+ dw^-+\frac{k^2\, \Delta_\theta(\theta) \sin^2\theta}{y^2 \rho_-(\theta)^2}dy^2+\frac{\rho_-(\theta)^2}{\Delta_\theta(\theta)}d\theta^2\nonumber +...\,,\nonumber
\eea
where 
\bea
\rho_-(\theta)^2=r_-^2+a^2\cos^2\theta\,,\qquad k= 2 a /\tilde{\delta}\,.
\eea
In this case we find the parameters in the transformation become 
\bea
\tilde{\alpha}=\frac{\Delta'(r_-)}{2 \,a\Xi}\,,\qquad \tilde{\beta}=0\,,\qquad\tilde{\gamma}=\frac{(\Omega_+ +\Omega_-)}{(\Omega_- -\Omega_+)}\tilde{ \alpha}
\,,\qquad \tilde{\delta}=-\Omega_- \, (\tilde{\alpha}+\tilde{\gamma})\,,
\eea
or 
\bea
\tilde{\alpha}={2\pi T_R}\,,\qquad \tilde{\beta}=0\,,\qquad \tilde{\gamma}=2\pi T_L
\,,\qquad \tilde{\delta}=-\frac{2\pi a\Xi \, (T_L+T_R)}{(a^2+r_-^2)}\,.
\eea
To proceed with the interpretation of the right and left-temperatures arising from the CFT we compute (\ref{temRL}) that brings them to the form
\bea\label{AdsTempIn}
T_R=\frac{\Delta'(r_-)}{4\pi \,a\Xi}\,,\qquad T_L=\frac{(\Omega_++\Omega_-)}{(\Omega_- -\Omega_+)}\frac{\Delta'(r_-)}{4\pi \,a\Xi }\,.
\eea
%

The general result (\ref{AdsTempIn}) for the right and left-temperatures arising from the CFT for the AdS-Kerr black holes can be combined with the Cardy entropy formula 
\begin{equation}
S_{\pm}=\frac{\pi^2}{3}\left(c_L T_{L}\pm c_R T_R \right)\,,
\end{equation}
using (\ref{AdsTempIn}) and the entropies (\ref{thermodynamicsAdS}), that for Kerr black holes the left and right CFTs have identical central charges
\bea\label{centralAdsKerr}
c_L=c_R=-\frac{6 a  }{\tilde{\delta}}=\frac{ 6 a (r_-^2-r_+^2)}{\Delta'(r_-)}\,.
\eea
%

\section{Discussion}
\label{sec:discussion}

The hidden conformal symmetries of black holes are of particular interest to explain the leading black hole microstate degeneracy.
Because many of the most intriguing results along these lines have been found in the context of asymptotically flat black holes, the extension of the soft hair approach to include spacetimes with cosmological constant is both natural and important. 

Our investigations of the soft hair has allowed us to establish a set of infinitesimal diffeomorphisms that act non-trivially on the horizon of generic Kerr-AdS black holes. Amongst these are a left-right pair of Virasoro algebras with associated charges that reproduce the Bekenstein-Hawking entropy for AdS black holes.

The fact that there are generic near-AdS$_3$ regions exhibiting the conformal structure of the black hole geometry is in general not sufficient to determine the conformal coordinates relevant to the application of this formalism. We argued that the Noether charge interpretation of the entropy in \cite{Wald:1993nt} gives the necessary input to uniquely fix the relevant conformal coordinates. Adding a boundary counterterm, justified to achieve integrability, lead to well-defined central charges $c_L = c_R$. 
We do not herein prove uniqueness of the counterterm or attempt to tackle other difficult problems related to the characterizing diffeomorphisms or charges. 

It is worth emphasizing that the choice of counterterm considered in our paper - that generalizes the Wald-Zoupas counterterm for asymptotically AdS space-times - gives a result that is compatible with all other results in the literature for Kerr-AdS black holes, see e.g. \cite{Hartman:2008pb,Lu:2008jk}. In the case of asymptotically flat rotating black holes and electrically charged black holes, the contributions from the counterterm reduce to the same ones from the Wald-Zoupas counterterm \cite{Haco:2018ske,Haco:2019ggi}.

A connection between the generic Kerr-AdS black hole results in this paper and previous results involving their extremal counterparts indicates a close link between the different computational approaches. By providing a first law of thermodynamics we also made a step in giving further evidence of the robustness of the  soft hair formalism in reproducing the Bekenstein-Hawking entropy.

\section*{Acknowledgements}

We are grateful to Andrew Strominger for useful conversations. We would like to thank the Mitchell Family Foundation for hospitality in 2018 at the Brinsop Court in England and 2019 Cook's Branch workshop and for continuing support. Part of this work was conducted at the `Fourth USU Strings and Black Holes Workshop', which was supported by the Department of Physics and the DGCAMP group at Utah State University. We would also like to thank the Centro de Ciencias de Benasque Pedro Pascual for the hospitality where some of the research was carried out. The work of MJR is partially supported through the NSF grant PHY-1707571, SEV-2016-0597 and PGC2018-095976-B-C21 from MCIU/AEI/FEDER, UE. MJP is supported by an STFC consolidated grant ST/L000415/1, String Theory, Gauge Theory and Duality.

\appendix


\section{Appendix: Conformal Coordinates for Kerr Black Holes}
\label{app:Kerr}

The metric of the four dimensional Kerr black hole,  satisfying $R_{\mu\nu}=0$ is given by
\bea
ds^2&=&\rho^2\left(\frac{dr^2}{\Delta}+{d\theta^2}\right)+\frac{\sin^2\theta}{\rho^2}\left(a\,dt-(r^2+a^2)\,d\phi\right)^2-\frac{\Delta}{\rho^2}\left(dt-a\sin^2\theta d\phi\right)^2\,,\nonumber\\
\rho^2&=&r^2+a^2\cos^2\theta\,\qquad \Delta=r^2+a^2-2Mr\,,
\eea
Note that the metric is asymtotically flat. The outer and inner event horizons are located at $\Delta(r_{\pm})=0$. The physical parameters are given by
\bea \label{KerrEntropy}
T_{\pm}&=&\frac{r_\pm^2-a^2}{4\pi r_\pm(r_\pm^2+a^2)}\,,\qquad S_{\pm}={\pi (r_{\pm}^2+a^2)}\,,\nonumber\\
J&=&M a\,,\qquad \Omega_{\pm}=\frac{a}{r_\pm^2+a^2}\,.
\eea
Employing the transformation (\ref{eq:coordtransf}) with (\ref{trtl1}), as well as 
$c = 1/\Delta'(r_+)$ in the function $R^2(r) \sim c \,(r- r_+)$ close to the event horizon of the black hole, the leading order of the black hole metric around the bifurcation surface becomes
\bea
ds^2= \frac{4\,\rho_+(\theta)^2}{y^2}dw^+ dw^-+\frac{16 J^2 \sin^2\theta}{y^2 \rho_+(\theta)^2}dy^2+\rho_+(\theta)^2d\theta^2+...\,,
\eea
where $\rho_+(\theta)^2=r_+^2+a^2\cos^2\theta$, we find the parameters
\bea
\alpha=\frac{\Delta'(r_+)}{2 \,a}\,,\qquad \beta=0\,,\qquad\gamma=\frac{(\Omega_++\Omega_-)\Delta'(r_+)}{2 \,a(\Omega_- -\Omega_+)}\,,\qquad \delta=-\frac{\Omega_- \Delta'(r_+)}{(r_+^2+a^2) (\Omega_- -\Omega_+)}
\eea
that can be written more explicitly by
\bea
\alpha=\frac{r_+-r_-}{2 \,a}\,,\qquad \beta=0\,,\qquad\gamma=\frac{r_++r_-}{2\,a}\,,\qquad \delta=-\frac{1}{2 M}
\eea
Now, we can define the left and right-moving temperatures via (\ref{temRL})
\bea \label{KerrTempRL}
T_R=\frac{r_+-r_-}{4\pi \,a}\,,\qquad T_L=\frac{r_++r_-}{4 \pi \,a}\,.
\eea
This agrees perfectly with the results in \cite{Haco:2018ske}. Moreover, at extremality the right temperature vanishes, while the left temperature goes to the result obtained in \cite{Guica:2008mu} in the extremal limit (where $r_+=r_-$):
\bea 
T_R=0\,,\qquad T_L=\frac{1}{2 \pi} \qquad \text{(at extremality)}\,.
\eea
The structure we find is generic (in particular it is valid for extremal, and non-extremal Kerr black holes). Together with the fact that the Cardy entropy formula is defined by
\begin{equation}
S_{\pm}=\frac{\pi^2}{3}\left(c_L T_{L}\pm c_R T_R \right)\,,
\end{equation}
we can show, using (\ref{KerrTempRL}) and the entropies (\ref{KerrEntropy}), that for Kerr black holes the left and right CFTs have identical central charges
\bea
c_L=c_R=\frac{ 6 a (r_+^2-r_-^2)}{\Delta'(r_+)}= 12 J\,.
\eea
%


\end{document}